\documentclass[letterpaper,twocolumn]{IEEEtran}
\usepackage{amsfonts}
\usepackage{amsmath,cases,amssymb}
\usepackage[dvips]{graphicx}
\usepackage{cite,color}
\usepackage{color}
\usepackage{subcaption}
\usepackage{epsfig}
\usepackage{epstopdf}
\usepackage{url}
\usepackage{algorithm}
\usepackage{algorithmicx, algpseudocode}
\usepackage{color}
\usepackage{multirow}
\usepackage{mathtools}
\usepackage{balance}
\newcommand{\rank}{{\sf rank}}

\newcommand{\clL}{{\cal L}}

\newcommand{\ds}{\displaystyle}

\newcommand{\clF}{{\cal F}}

\newcommand{\clN}{{\cal N}}
\newcommand{\clG}{{\cal G}}

\newcommand{\Tr}{\mbox{Trace}}

\newcommand{\bgeqn}{\begin{equation}}
\newcommand{\edeqn}{\end{equation}}

\newcommand{\beqa}{\begin{eqnarray}}
\newcommand{\eeqa}{\end{eqnarray}}
\newcommand{\beqas}{\begin{eqnarray*}}
\newcommand{\eeqas}{\end{eqnarray*}}

\newcommand{\clT}{{\cal T}}
\newcommand{\clP}{{\cal P}}
\newcommand{\clV}{{\cal V}}
\newcommand{\clW}{{\cal W}}
\newcommand{\clR}{{\cal R}}
\begin{document}
\title{Model Predictive Control for  Smart Grids with Multiple Electric-Vehicle Charging Stations}
\author{Y. Shi, H. D. Tuan, A. V. Savkin, T. Q. Duong and H. V.  Poor
\thanks{Ye Shi and Hoang D. Tuan are with the School of Electrical and Data Engineering, University of Technology Sydney, Broadway, NSW 2007, Australia (email: Ye.Shi@student.uts.edu.au, Tuan.Hoang@uts.edu.au);
Andrey V. Savkin is with the School of Electrical Engineering and Telecommunications, The University of New South Wales
Sydney, NSW 2052, Australia (email: a.savkin@unsw.edu.au); Trung Q. Duong is with Queen's University Belfast, Belfast BT7 1NN, UK  (email: trung.q.duong@qub.ac.uk); H. Vincent Poor is with the Department of Electrical Engineering, Princeton University, Princeton, NJ 08544, USA (e-mail: poor@princeton.edu)}
}
\date{}
\maketitle
\begin{abstract} Next-generation  power grids  will likely enable  concurrent
service for residences and plug-in electric vehicles (PEVs). While the residence power demand profile is
known and thus can be considered inelastic, the PEVs' power demand is only known after random PEVs' arrivals.
PEV charging scheduling aims
at minimizing the potential impact of the massive integration of PEVs into  power grids
to save  service costs to customers while power control aims at minimizing the cost
of power generation subject to operating constraints and meeting demand.
The present paper develops a model predictive
control (MPC)-based approach to address the joint PEV charging scheduling and power control to minimize both
PEV charging cost and energy generation cost in meeting both residence and PEV power demands.
 Unlike in related works,
no assumptions are made about  the probability distribution of PEVs' arrivals,  the known
PEVs' future demand, or the unlimited charging capacity of PEVs.
 The proposed approach is shown to achieve a globally optimal solution.
Numerical results for IEEE benchmark power grids serving Tesla Model S PEVs  show the merit of this approach.
\end{abstract}
\section{Introduction}
Electrical vehicles (EVs) have emerged  as a promising solution to resolve both the economic and environmental concerns in
the transportation industry  \cite{Tuetal16}. Using a smart power grid in concurrently serving residences
and charging EVs constitutes one of the most important applications of the  smart grid technology.
However, the massive integration
of plug-in EVs (PEVs) into the grid
causes many potential impacts such as voltage deviation, increased load variations and power loss of
the grid \cite{TBZ16}, which  requires different strategies for load shifting and
energy trading and storage in the grid \cite{Waetal14,YZP14,Waetal16,Laetal16}. The
main difficulty in scheduling of PEV charging   to manage the cost and impact of PEV integration  is that individual PEVs randomly arrive for
charging with their individual demands on charging load and deadlines, which cannot be known before hand. In other words,
the future charging demand of PEVs cannot be known a priori. Many existing works consider a simple smart grid with a single
charging station (CS) to exclusively serve PEVs. For instance, \cite{LZ12} sets no charging deadlines for PEVs, whose
arrival process follows a probability distribution, while \cite{MG15} assumes that the future load demand is
perfectly known a priori. The future load demand is also assumed to be known in \cite{Xietal16} as all PEVs are assumed to arrive at the same time
with no charging deadline. It is assumed in \cite{TZ17} that only statistics of demand are known but the PEVs can be fully charged
in a single time slot \cite[(30)]{TZ17}. It should be realized that serving PEVs is typically considered during a $12$-hour
time period (for instance from 8:00 pm to 8:00 am), where the integration of a massive number of PEVs has a sizable
effect on the power grid, and as such, the length of a time slot is rationally set by $30$ minutes or
one hour. In other words, the charging scheduling should be considered over a finite horizon of $12$-$24$ time slots, but not
over an infinite horizon as considered in \cite{KKC17}.
Due to their physical limitations, PEVs are rarely able to be fully charged just during a single time slot.

In this paper, we consider  joint PEV charging scheduling and power control to save service costs for
PEVs and the power generation costs in meeting both residential and PEV power demands.
Such a problem  was considered in \cite{CTQ14} but only a small number of PEVs
and with  each CS serving only one PEV, whose power demand
is very small compared with the inelastic demand, so that
the integration of PEVs into the grid has almost no effect on the grid.
Note that the  optimal power flow problems posed in \cite{CTQ14} cannot be solved exactly by
semi-definite programming relaxation (SDR) \cite{Yeetal17}. Therefore, it is not known if the objective
in PEVs charging scheduling is convex and as such, it is not known if its proposed valley-filling
solution is optimal. Larger PEVs' penetration in a few  CSs was considered in \cite{FRR15,ZKG17} under the assumption
of known
arrival and departure times of PEVs. In the present paper,
we are interested in more practical scenarios of a massive number of PEVs arriving
randomly at different CSs. No assumption on the probability distribution of their arrival is made, so the conventional
model predictive control (MPC) \cite{Ca04,M16} is not applicable. Our contribution is to develop a novel MPC-based
approach to address this problem.

The rest of the paper is structured as follows. Section II is devoted to the system
modeling for this problem and  analyzing its computational challenges. An online computational solution
using the proposed MPC-based approach is developed in Section III. An off-line computational
solution is considered in Section IV, which is then compared with the online computational solution in Section V
to show the optimality of the later. Section VI concludes the paper.

{\it Notation.} The notation used in this paper is standard. Particularly, $j$ is the imaginary unit,
$X^H$ is Hermitian transpose  of a vector/matrix $X$, $M\succeq 0$ for a Hermitian symmetric matrix $M$ means that
it is positive semi-definite, $\rank(M)$ and $\Tr(M)$ are the rank and trace of a matrix $M$, respectively. $\Re (\cdot)$ and $\Im (\cdot)$ are the real and imaginary parts of a complex quantity, and
 $a\leq b$ for two complex numbers $a$ and $b$ is componentwise understood, i.e. $\Re(a)\leq \Re(b)$ and $\Im(a)\leq \Im(b)$.
The cardinality of a set ${\cal C}$ is denoted by $|{\cal C}|$.
\section{Problem statement and computational challenges}
\begin{figure}[h]
\centering
\includegraphics[width=0.9 \columnwidth]{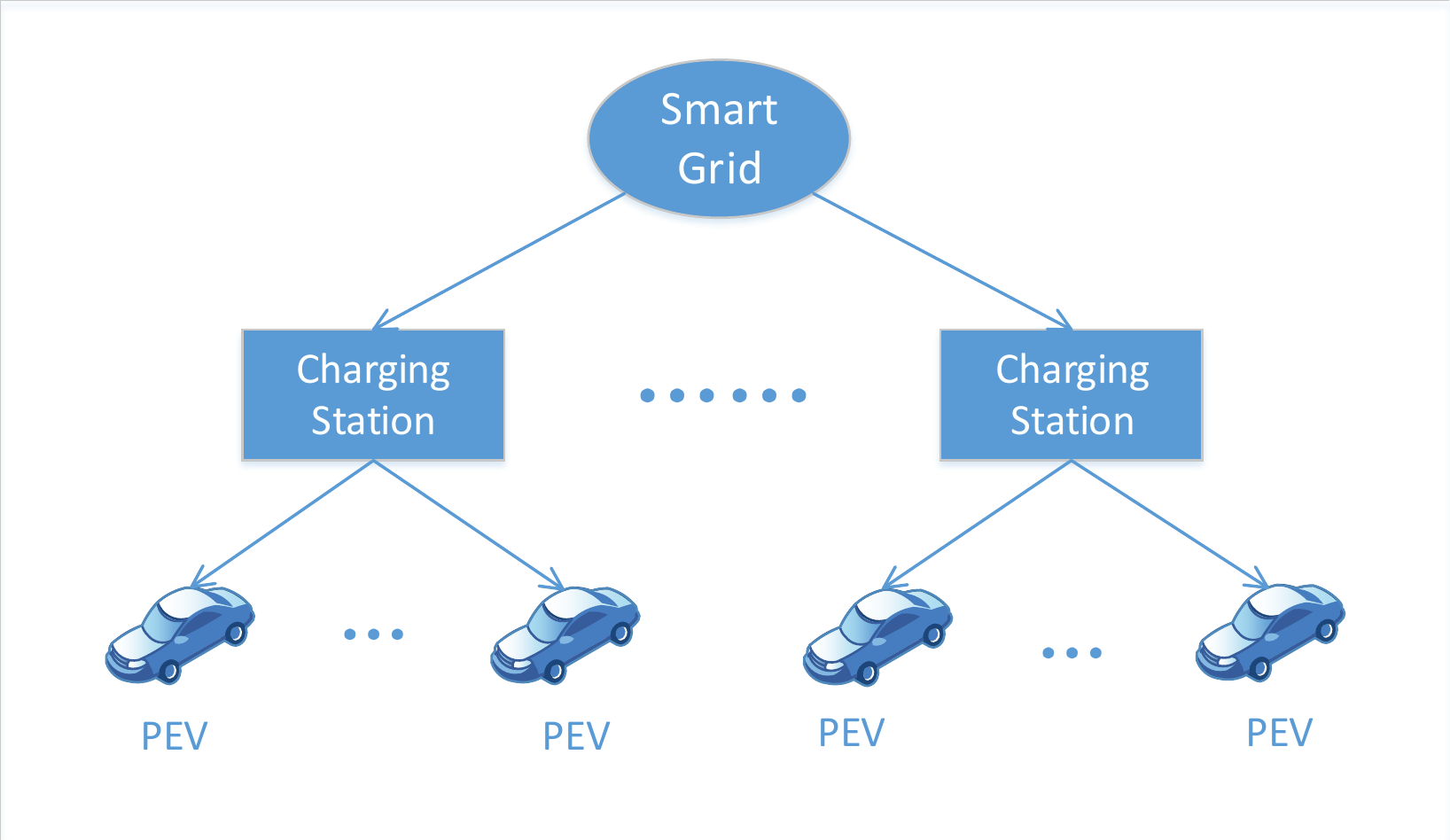}
\caption{A simplified connection between PEVs and CSs in the smart grid}
\label{PEV_CS}
\end{figure}

Consider an eletricity grid with a set of buses ${\cal N} := \{1, 2,..., N\}$ connected
through a set of flow lines ${\cal L}\subseteq {\cal N}\times {\cal N}$, i.e. bus $k$ is connected to bus $m$ if and only if $(k,m)\in {\cal L}$. Accordingly, $\clN(k)$ is the set of other buses connected to bus $k$.
There is a subset ${\cal G}\subseteq {\cal N}$, whose elements are connected to distributed generators (DGs).
Any bus $k\notin {\cal G}$ is thus not connected to DGs. Any bus $k\in {\cal G}$ also has a function
to serve PEVs and in what follow is also referred to  CS $k$. By defining $M=|{\cal G}|$, there are $M$ CSs in the grid.
Denote by ${\cal H}_k$ the set of those PEVs that arrive at CS $k$. Accordingly, $k_n$ is the $n$-th PEV that arrives
at  CS $k$. Figure \ref{PEV_CS} presents a block diagram illustrating  PEV integration into the grid.

The serving time period of the grid is divided into $T$ time slots of length $\delta_t$, which usually varies from $30$ minutes to an hour. Under the definition $\clT:=\{1, 2,\dots, T\}$,
PEV $k_n$ arrives at $t_{a,k_n}\in \clT$ and needs to depart at $t_{k_n,d}\in\clT$. The constraint
\begin{equation}\label{time1}
t_{k_n,d}-t_{k_n,a}\leq T_{k_n},
\end{equation}
expresses the PEV $k_n$'s time demand.
Suppose that $C_{k_n}$ and $s_{k_n}^0$ are the battery capacity and initial state of charge (SOC) of PEV $k_n$.
It must be fully charged by the departure time $t_{k_n,d}$, i.e.
\begin{eqnarray}\label{HEV1h}
\ds\sum_{t'=t_{k_n,a}}^{t_{k_n,d}} u_hP_{k_n}(t') =  C_{k_n}(1 - s_{k_n}^{0}),
\end{eqnarray}
where $u_h$ is the charging efficiency of the battery and $P_{k_{n}}(t')$ is a decision variable representing the power charging rate of PEV $k_n\in {\cal H}_k$  at time $t'$.

Due to the limited capacity of the hardware, the following constraint must be imposed:
\begin{equation}\label{HEV1i}
0 \leq P_{k_n}(t') \leq {\overline P}_{k_n}, t_{k_n,a}\leq t'\leq t_{k_n,d},
\end{equation}
for a given ${\overline P}_{k,n}$.
For ease of presentation, we set
\begin{equation}\label{soca}
P_{k_n}(t')=0,  t'\notin [t_{k_n,a}, t_{k_n,b}].
\end{equation}
Let $y_{km}\in\mathbb{C}$ be the admittance of line $(k,m)$. The current $I_k(t')$ at node $k\in \clN$ is
$I_{k}(t') = \sum_{m\in \clN(k)}I_{km}(t') = \sum_{m\in \clN(k)} y_{km} (V_k(t') - V_m(t'))$,
where $V_k(t')$ is the complex voltage at bus $k$ during the time slot $t'$.
For $k\in \clG$, the total supply and demand energy is balanced as
$V_k(t')(I_k(t'))^* =  (P_{g_{k}}(t') - P_{l_{k}}(t') -\sum_{n\in {\cal H}_k}P_{k_{n}}(t'))+ j(Q_{g_{k}}(t') - Q_{l_{k}}(t'))$,
where $P_{g_{k}}(t')$ and $Q_{g_{k}}(t')$ are the real
and reactive powers generated by DG $k$, and  $P_{l_{k}}(t')$ and $Q_{l_{k}}(t')$ are respectively
known real  and reactive price-inelastic demands
at bus $k$ to express the residential power demand. By using
the last two equations, we obtain
\begin{eqnarray}\label{HEV1b}
 V_k(t')(\sum_{m\in \clN(k)} y_{km} (V_k(t') - V_m(t')))^* =  (P_{g_{k}}(t') \nonumber \\
 - P_{l_{k}}(t')  -\sum_{n\in {\cal H}_k}P_{k_{n}}(t')) + j(Q_{g_{k}}(t') - Q_{l_{k}}(t')), k\in \clG.
\end{eqnarray}
Similarly,
\begin{eqnarray}\label{HEV1c}
V_k(t')(\sum_{m\in \clN(k)} y_{km} (V_k(t') - V_m(t')))^* = \nonumber \\
- P_{l_{k}}(t')  - j Q_{l_{k}}(t'),  k\notin  {\cal G}.
\end{eqnarray}
The next constraints relate to the acceptable range of generated power by the DGs:
\begin{eqnarray}
{\underline P}_{g_k} \leq P_{g_k}(t') \leq {\overline P}_{g_k}\ \&\
{\underline Q}_{g_k} \leq Q_{g_k}(t') \leq {\overline Q}_{g_k}, \  k\in {\cal G}, \label{HEV1e}
\end{eqnarray}
where ${\underline P}_{g_k}$, ${\underline Q}_{g_k}$ and ${\overline P}_{g_k}$, ${\overline Q}_{g_k}$ are the the lower limit and upper limit of the real generated  and reactive generated powers, respectively.

The constraints of voltage  are
\begin{eqnarray}
{\underline V}_k \leq |V_k(t')| \leq {\overline V}_k, \quad  k\in {\cal N},\label{HEV1f}\\
|\mbox{arg}(V_k(t'))-\mbox{arg}(V_m(t'))| \leq \theta_{km}^{\max}, (k,m)\in\clL, t'\in\clT, \label{HEV1g}
\end{eqnarray}
where ${\underline V}_k$ and ${\overline V}_k$ are the lower limit and upper limit of the voltage amplitude,
while $\theta^{\max}_{k,m}$ are given to express the voltage phase balance.

The problem of interest  is to minimize both the energy cost to DGs and charging cost for PEVs. Thus,
by defining $V(t')=(V_1(t'),\dots, V_N(t'))$, $\clV=\{V(t')\}_{t'\in\clT}$,
and $P_g(t')=(P_{g_1}(t'),\dots, P_{g_M}(t'))$,
$Q_g(t')=(Q_{g_1}(t'),\dots, Q_{g_M}(t'))$,
$R(t')=\{P_g(t'), Q_g(t')\}$,  $\clR=\{R(t')\}_{t'\in\clT}$,
and $\clP^{PEV}=\{P^{PEV}(t')\}_{t'\in\clT}$,
$P^{PEV}(t')=\{P_{k_n}(t')\}_{k_n\in{\cal H}_k, k=1,\dots, M}$, the objective
function is given by $F(\clR,\clP^{PEV})=\ds \sum_{t'\in \clT} \sum_{k\in {\cal G}}f(P_{{g_k}}(t')) + \ds \sum_{t'\in \clT} \sum_{k\in {\cal N}} \sum_{n\in {\cal H}_k} \beta_tP_{{k_n}}(t')$,
where $f(P_{{g_k}}(t'))$ is the cost function of real power generation  by DGs, which is  linear or quadratic in $P_{{g_k}}(t')$, and $\beta_t$ is the known PEV charging price during the time interval $(t',t'+1]$.

The joint PEV charging scheduling and power control is then mathematically formulated as
\begin{equation}\label{HEV1}
\ds\min_{\clV, \clR,\clP^{PEV}}\ F(\clR,\clP^{PEV})\quad
\mbox{s.t.}\quad (\ref{HEV1h})-(\ref{HEV1g}).
\end{equation}
The above problem (\ref{HEV1}) is very computationally challenging because the quadratic equality constraints
(\ref{HEV1b}) and (\ref{HEV1c}) and nonlinear inequality constraints (\ref{HEV1f}) and (\ref{HEV1g}) constitute
 nonconvex constraints. Moreover, the arrival time $t_{k_n,a}$ of each individual PEV $k_n$, its charging
demand and its departure time $t_{k_n,d}$ are unknown.
\section{Model predictive control (MPC)-based computational solution}
Considering $(R(t'), P^{PEV}(t'))$ and $V(t')$ as the plant state and control, respectively,
equations (\ref{H EV1b}), (\ref{HEV1c}), and (\ref{HEV1e}) provide
 state behavioral  equations \cite{PW} with the end constraint (\ref{HEV1h}) while equations (\ref{HEV1f})
and (\ref{HEV1g}) provide control constraints.  On the surface,
(\ref{HEV1}) appears to be a control problem over the finite horizon $[1,T]$.  However, all equations in (\ref{HEV1}) are unpredictable beforehand, preventing the application of conventional model predictive control \cite{Ca04,M16}. We now follow the
idea of \cite{Tuetal15} to address (\ref{HEV1}).

At each time $t$ denote by $C(t)$ the set of PEVs that need to be charged. For each $k_n\in C(t)$, let
$\clP_{k_n}(t)$ be its remaining demand for charging by the departure time $t_{k_n,d}$. Define
\begin{equation}\label{hori1}
\Psi(t)=\max_{k_n\in C(t)} t_{k_n,d}.
\end{equation}
At time $t$ we solve the following optimal power flow  (OPF) problem over
the prediction horizon $[t,\Psi(t)]$ but then take only
$V(t),  P_{k_n}(t),  R(t)$ for online updating solution of (\ref{HEV1}):
\allowdisplaybreaks[4]
\begin{subequations}\label{HOR1}
\begin{eqnarray}
\ds\min_{V(t'), R(t'), P_{k_n}(t'), t'\in [t,\Psi(t)], k_n\in C(t)} F_{[t,\Psi(t)]}&& \label{HOR1a}\\
\mbox{s.t.}\quad (\ref{HEV1i})-(\ref{soca}), (\ref{HEV1c})-(\ref{HEV1g})\quad \mbox{for}\ t'\in [t,\Psi(t)],
 \label{HOR1g}\\
 V_k(t')(\sum_{m\in \clN(k)} y_{km} (V_k(t') - V_m(t')))^* = \nonumber \\
(P_{g_{k}}(t') - P_{l_{k}}(t') -\sum_{k_n\in C(t)}P_{k_{n}}(t'))
\nonumber\\ + j(Q_{g_{k}}(t') - Q_{l_{k}}(t')),  \quad  (t',k)\in [t,\Psi(t)]\times \clG,  \label{HOR1b}  \\
\ds\sum_{t'=t}^{t_{k_n,d}} u_hP_{k_n}(t') =\clP_{k_n}(t),\label{HOR1h}
\end{eqnarray}
\end{subequations}
with $F_{[t,\Psi(t)]}:=\ds \sum_{t'=t}^{\Psi(t)}
\sum_{k\in {\cal G}}f(P_{{g_k}}(t')) + \ds \sum_{t'=t}^{\Psi(t)} \sum_{k_n\in C(t)}\beta_tP_{{k_n}}(t') $.
One can notice that (\ref{HOR1}) includes only what is known at the present time $t$. Of course, (\ref{HOR1}) is a still difficult nonconvex optimization and in the end we need only its solution
at $t$, so we propose the following approach in tackling its solution at $t$.

Define the Hermitian symmetric matrix  $W(t') = V(t')V^H(t') \in \mathbb{C}^{N\times N}$,
which must satisfy $ W(t')\succeq 0$ and $\rank(W(t'))=1$.
By replacing  $W_{km}(t')=V_k(t')V^*_m(t')$,  $(k,m)\in\clN\times\clN$,  in (\ref{HOR1b})-(\ref{HOR1g}),
we reformulate (\ref{HOR1}) to the following optimization problem in matrices $W(t')\in\mathbb{C}^{N\times N}$,
$t'\in [t,\Psi(t)]$:
\allowdisplaybreaks[4]
\begin{subequations}\label{rHOR1}
\begin{eqnarray}
\ds\min_{W(t'), R(t'), P_{k_n}(t'), t'\in [t,\Psi(t)], k_n\in C(t)} F_{[t,\Psi(t)]} \label{rHOR1a}\\
\mbox{s.t.}\quad   (\ref{HEV1i})-(\ref{soca}), (\ref{HEV1e})\quad\mbox{for}\ t'\in [t,\Psi(t)],\label{rHOR1d}\\
\ds\sum_{m\in \clN(k)} (W_{kk}(t') - W_{km}(t'))y_{km}^*  = (P_{g_{k}}(t') - P_{l_{k}}(t')\nonumber \\
 -\sum_{k_n\in C(t)}P_{k_{n}}(t'))
 + j(Q_{g_{k}}(t') - Q_{l_{k}}(t')),  \quad  k\in \clG,  \label{rHOR1b}  \\
\sum_{m\in \clN(k)}(W_{kk}(t') - W_{km}(t'))y_{km}^* =   \nonumber\\
 - P_{l_{k}}(t')  - j Q_{l_{k}}(t'), k\notin  {\cal G}, \label{rHOR1c}\\
{\underline V}_k^2 \leq W_{kk}(t') \leq {\overline V}_k^2, \quad  k\in {\cal N}, \label{rHOR1f}\\
\Im(W_{km}(t'))\leq \Re(W_{km}(t'))\tan(\theta_{km}^{max}), (k,m)\in\clL,  \label{rHOR1g}\\
W(t')\succeq 0,  \label{rHOR1k}\\
\mbox{rank}(W(t'))=1.\label{rHOR1l}
\end{eqnarray}
\end{subequations}
Instead of (\ref{rHOR1}), which is difficult due to multiple nonconvex  matrix rank-one constraints in
(\ref{rHOR1k}), we solve its semi-definite relaxation (SDR)
\begin{eqnarray}
\ds\min_{W(t'), R(t'), P_{k_n}(t')} F_{[t,\Psi(t)]}\quad
 \mbox{s.t.}\ (\ref{rHOR1d})-(\ref{rHOR1k}).\label{HORr}
\end{eqnarray}
Suppose that $\hat{W}(t')$ and $(\hat{R}(t'),\hat{P}_{k_n}(t'))$, $t'\in [t,\Psi(t)]$ are the optimal solution of (\ref{HORr}). If
$\mbox{rank}(\hat{W}(t'))\equiv 1$, $t'\in [t,\Psi(t)]$, then $\hat{V}(t')$ such that $\hat{W}(t')=\hat{V}(t')\hat{V}^H(t')$
together with $\hat{R}(t')$ and  $\hat{P}_{k_n}(t')$ constitute
the optimal solution of the nonconvex optimization problem (\ref{HOR1}). Otherwise, we consider the following problem:
\begin{subequations}\label{rHOR2}
\begin{eqnarray}
\ds\min_{W(t), R(t)}\ F(P_g(t))):=\ds \sum_{k\in {\cal G}}f(P_{{g_k}}(t))&& \label{rHOR2a}\\
\mbox{s.t.}\quad  (\ref{HEV1i})-(\ref{soca}), (\ref{HEV1e}), (\ref{rHOR1c})-(\ref{rHOR1k})\quad\mbox{for}\ t'=t,\label{rHOR2k}\\
\ds\sum_{m\in \clN(k)} (W_{kk}(t) - W_{km}(t))y_{km}^*  =(P_{g_{k}}(t) - P_{l_{k}}(t)\nonumber \\
 -\sum_{k_n\in C(t)}\hat{P}_{k_{n}}(t))
+ j(Q_{g_{k}}(t) - Q_{l_{k}}(t)),  \quad  k\in \clG,  \label{rHOR2b} \\
\mbox{rank}(W(t))=1.\label{rHOR2l}
\end{eqnarray}
\end{subequations}
Note that in contrast to (\ref{rHOR1}) involving $\Psi(t)-t$ matrix variables $W(t')$, $t'\in [t,\Psi(t)]$ and also
variables $P_{k_n}(t')$, $k_n\in C(t)$ and $t'\in [t,\Psi(t)]$, there is only single matrix variable $W(t)$ in (\ref{rHOR2}).
The power generation variable $R(t)$ in (\ref{rHOR2}) is latent as it is inferred from $W(t)$ in equation (\ref{rHOR2b}).

Following our previous works \cite{Phetal12,STST15,Naetal17,Yeetal17,STA17}, a nonsmooth optimization algorithm (NOA) is proposed to deal with the discontinuous matrix rank-one constraint (\ref{rHOR2l}) in the optimization problem (\ref{rHOR2}).
Under condition (\ref{rHOR1k}) in (\ref{rHOR2k}),
\[
\Tr(W(t))-\lambda_{\max}(W(t))\geq 0,
\]
where $\lambda_{\max}(W(t))$ stands for the maximal eigenvalue of $W(t)$.
The discontinuous matrix rank-one constraint (\ref{rHOR2l}) is then equivalently expressed by the following
continuous  spectral constraint:
\begin{equation}\label{rHEV3ie}
\Tr(W(t))-\lambda_{\max}(W(t)) = 0,
\end{equation}
because it means that $W(t)$ has only one nonzero eigenvalue. Thus the quantity $\Tr(W(t))-\lambda_{\max}(W(t))$
expresses the degree of the matrix rank-one constraint satisfaction  (\ref{rHEV3ie}), which is incorporated into
 the objective (\ref{rHOR2a}), leading to the following penalized optimization problem:
\begin{eqnarray}\label{rW3}
\ds\min_{W(t), R(t)} F_{\mu}(W(t),P_g(t)):=F(P_g(t)) + \nonumber \\ \mu(\Tr(W(t))-\lambda_{\max}(W(t)))
\quad \mbox{s.t.}\quad (\ref{rHOR2b})-(\ref{rHOR2k}),
\end{eqnarray}
where $\mu>0$ is a penalty parameter. The above penalized optimization is
exact because the constraint (\ref{rHOR2k}) can be satisfied by a minimizer of (\ref{rW3}) with a finite value of $\mu$.
On the other hand, any  $W(t)$ feasible for (\ref{rW3}) is also feasible
for (\ref{rHOR2}), implying that the optimal value of (\ref{rW3}) for any $\mu > 0$ is upper bounded by the optimal value
of (\ref{rHOR2}).

For any $W^{(\kappa)}(t)$ feasible for the convex constraints (\ref{rHOR2b})-(\ref{rHOR2k}), let
$w^{(\kappa)}_{\max}(t)$ be the normalized eigenvector
corresponding to the eigenvalue $\lambda_{\max}(W^{(\kappa)}(t))$. Then
\begin{eqnarray}\label{rW4}
\lambda_{\max}(W(t))&=&\max_{||w||^2=1}w^HW(t)w\nonumber\\
&\geq&(w_{\max}^{(\kappa)}(t))^H W(t) w_{\max}^{(\kappa)}(t),
\end{eqnarray}
i.e. the function $\lambda_{\max}(W(t))$ is lower bounded by the linear function $(w_{\max}^{(\kappa)}(t))^H W(t) w_{\max}^{(\kappa)}(t)$. Accordingly, the following semi-definite program (SDP) provides an upper bound for the nonconvex optimization problem (\ref{rW3}):
\begin{eqnarray}\label{rW5}
\ds\min_{W(t), R(t)} F_{\mu}^{(\kappa)}(W(t),R(t)):=F(P_g(t))+\mu (\Tr(W(t))
\nonumber \\ -(w_{\max}^{(\kappa)}(t))^H W(t) w_{\max}^{(\kappa)}(t))\quad
\mbox{s.t.}\quad (\ref{rHOR2b})-(\ref{rHOR2k}),
\end{eqnarray}
because $F_{\mu}^{(\kappa)}(W(t), R(t))\geq F_{\mu}(W(t), R(t))$ according to (\ref{rW4}).

Suppose that $(W^{(\kappa+1)}(t), R^{(\kappa+1)}(t))$ is the optimal solution of SDP (\ref{rW5}).
Since $(W^{(\kappa)}(t), R^{(\kappa)}(t))$ is also
feasible for (\ref{rW5}), it is true that
$F_{\mu}(W^{(\kappa)}(t), R^{(\kappa)}(t))=F_{\mu}^{(\kappa)}(W^{(\kappa)}(t), R^{(\kappa)}(t))
\geq F_{\mu}^{(\kappa)}(W^{(\kappa+1)}(t), R^{(\kappa)+1}(t))\geq F_{\mu}(W^{(\kappa+1)}(t), R^{(\kappa+1)}(t))$,
so $W^{(\kappa+1)}(t)$ is a better feasible point of (\ref{rW3}) than $W^{(\kappa)}(t)$.

In Nonsmooth Optimization Algorithm (NOA) \ref{alg1}  we  propose an iterative procedure, which is initialized by the solution
$\hat{W}(t)$ of SDR (\ref{HORr}) and generates a feasible point $W^{(\kappa+1)}(t)$ at the $\kappa$-th iteration for
$\kappa=0, 1, \dots$, as the optimal solution of SDP (\ref{rW3}).  As proved in \cite{Yeetal17}, this algorithm
converges at least to a local minimizer of (\ref{rW3}). Note that the procedure terminates at
$0\leq \Tr(W^{(\kappa)}(t))-\lambda_{\max}(W^{(\kappa)}(t))
\leq \Tr(W^{(\kappa)}(t))-(w_{\max}^{(\kappa)}(t))^H W^{(\kappa)}(t) w_{\max}^{(\kappa)}(t)
\leq \epsilon$, so the spectral constraint (\ref{rHEV3ie}) for the matrix rank-one is satisfied with the computational tolerance $\epsilon$.
\begin{algorithm}[!t]\caption{NOA \ref{alg1} for solving (\ref{rHOR2})}\label{alg1}
  \begin{algorithmic}[1]
  \State {\bf Set} $\kappa=0$ and $(W^{(0)}(t), R^{(0)}(t)) =(\hat{W}(t), \hat{R}(t))$.
\State {\bf Until} $ \Tr(W^{(\kappa)}(t))-(w_{\max}^{(\kappa)}(t))^H W^{(\kappa)}(t) w_{\max}^{(\kappa)}(t) \leq \epsilon$,
{\bf solve} (\ref{rW5}), to find the optimal solution $(W^{(\kappa+1)}(t), R^{(\kappa+1)}(t))$ and
{\bf reset} $\kappa+1\rightarrow \kappa$.
\State {\bf Accept} $(W^{(\kappa)}(t), R^{(\kappa)}(t))$  as the optimal solution of the nonconvex optimization problem (\ref{rHOR2}).
   \end{algorithmic}
\end{algorithm}
In summary, our proposed MPC-based computation for (\ref{HEV1}) is based on solving SDP (\ref{HORr}) for online
coordinating PEV charge $\hat{P}_{k_n}(t)$ and solving (\ref{rW3}) by NOA \ref{alg1} for online updating  the
generated voltage $\hat{V}(t)$ for the generated power $\hat{R}(t)$  by
\begin{equation}\label{Vt}
\hat{V}(t)=\sqrt{\lambda_{\max}(W^{(\kappa)})}w^{(\kappa)}_{\max}(t),
\end{equation}
whenever the solution $\hat{W}(t)$ of SDR (\ref{HORr}) is not of rank-one. If $\mbox{rank}(\hat{W}(t))=1$, it is obvious
that $\hat{V}(t)=\sqrt{\lambda_{\max}(\hat{W}(t))}\hat{w}_{\max}(t)$ with the normalized eigenvector $\hat{w}_{\max}(t)$
corresponding to $\lambda_{\max}(\hat{W}(t))$ is the optimal solution of (\ref{HOR1}), which is what we need.
\section{Lower bound by off-line optimization}
To investigate the optimality of the MPC-based online computation
proposed in the previous section, in this section we address an off-line computation for (\ref{HEV1}), which provides a lower bound for the optimal value of its online computation.
Under the additional definition $\clW=\{W(t')\}_{t'\in\clT}$, we reformulate (\ref{HEV1}) as
\begin{subequations}\label{HEV2}
\begin{eqnarray}
\ds\min_{\clW, \clR,\clP^{PEV}}\ F(\clR, \clR^{PEV})\quad\mbox{s.t.}&& \label{HEV2a}\\
 (\ref{HEV1h})-(\ref{soca}),   (\ref{HEV1e}), (\ref{rHOR1c})-(\ref{rHOR1k})\ \mbox{for}\ t'\in\clT,\label{HEV2d}\\
\sum_{m\in \clN(k)} (W_{kk}(t')-W_{km}(t'))y_{km}^*=(P_{g_{k}}(t') - P_{l_{k}}(t')\nonumber \\
- \sum_{n\in {\cal H}_k}P_{k_{n}}(t'))+j(Q_{g_{k}}(t') - Q_{l_{k}}(t')),  \ k\in\clG, \label{HEV2b}  \\
 \rank(W(t'))=1, t'\in \clT. \label{HEV2k}
\end{eqnarray}
\end{subequations}
First, we solve its SDR by dropping the matrix rank-one constraints in (\ref{HEV2k}):
\begin{equation}\label{HEV5}
\ds\min_{\clW, \clR,\clP^{PEV}}\ \clF_{\clT} \quad\mbox{s.t.} \quad (\ref{HEV2d})-(\ref{HEV2b}).
\end{equation}
Suppose that  $\hat{\clW}$ and $\hat{\clP}^{PEV}$ are the optimal solution of  SDP (\ref{HEV5}).
If $\mbox{rank}(\hat{W}(t))\equiv 1$, $t\in\clT$ then a global solution of the original problem (\ref{HEV1}) is found
as   $\hat{\clP}^{PEV}$, $\hat{\clR}$ and $\hat{\clV}$ and  with $\hat{V}(t)\hat{V}^H(t)=\hat{W}(t)$, $t\in\clT$.
However, such a matrix rank-one condition is rarely achieved. In what follows we propose two methods to address
the matrix rank-one constraints in
(\ref{HEV2k}).

Again, under  condition  (\ref{rHOR1k}) for $t'\in\clT$ in (\ref{HEV2d}), the rank-one constraints in (\ref{HEV2k}) are equivalently expressed by the single spectral constraint
$\sum_{t\in \clT}(\Tr(W(t))-\lambda_{\max}(W(t))) = 0$,
which is incorporated  into the objective function in (\ref{HEV2a}) for the following penalized function optimization:
\begin{eqnarray}\label{W3}
\ds\min_{\clW, \clR,\clP^{PEV}} \clF_{\clT}+ \mu\sum_{t\in \clT}(\Tr(W(t))-\lambda_{\max}(W(t))) \nonumber \\ \quad \mbox{s.t.}\quad (\ref{HEV2d})-(\ref{HEV2b}),
\end{eqnarray}
with  a penalty parameter $\mu>0$. Initialized by $\clW^{(0)}=\hat{\clW}$, the following SDP is solved
at $\kappa$-th iteration to generate $\clW^{(\kappa+1)}$ and $\clP^{PEV}$:
\begin{eqnarray}\label{W5}
\ds\min_{\clW, \clR,\clP^{PEV}} \clF_{\clT} +\mu\sum_{t\in \clT}(\Tr(W(t))-\nonumber \\(w_{\max}^{(\kappa)}(t))^H W(t) w_{\max}^{(\kappa)}(t)) \quad\mbox{s.t.}\quad (\ref{HEV2d})-(\ref{HEV2b}).
\end{eqnarray}
This computational procedure is summarized in NOA \ref{alg2}.
\begin{algorithm}[!t]\caption{ NOA \ref{alg2} for solving (\ref{HEV2})}\label{alg2}
  \begin{algorithmic}[1]
  \State {\bf Set} $\kappa = 0$ and $\clW^{(0)}=\hat{\clW}$.
  \State {\bf Until}
 $\ds\sum_{t\in \clT}(\Tr(W^{(\kappa)}(t))-(w_{\max}^{(\kappa)}(t))^H W^{(\kappa)}(t)w_{\max}^{(\kappa)}(t))$ $\leq$ $\epsilon$
  {\bf solve} (\ref{W5}) to generate $\clW^{(\kappa+1)}$, $\clR$ and $\clP^{PEV}$ and
  {\bf reset} $\kappa+1\rightarrow \kappa$.
  \State {\bf Accept} $\clW^{(\kappa)}$, $\clR$ and $\clP^{PEV}$  as the optimal solution of the nonconvex optimization problem (\ref{HEV1}).   \end{algorithmic}
\end{algorithm}

Alternatively, we propose the following scalable  algorithm for computing   (\ref{HEV2}).
By replacing $P_{k_n}(t)$ by $\hat{P}_{k_n}(t)$, which was found by solving from (\ref{HEV5}), in (\ref{HEV2}) at
every $t\in\clT$, we obtain the following optimization problem in $W(t)$ and $R(t)$ only:
\begin{subequations}\label{HEV3}
\begin{eqnarray}
\ds\min_{W(t), R(t)} \ds \sum_{k\in {\cal G}}f(P_{{g_k}})  \ \mbox{s.t.}\ (\ref{HEV2d})-(\ref{HEV2b})\
\mbox{for}\ t'=t,\\
 \rank(W(t))=1, \label{HEV3k}
\end{eqnarray}
\end{subequations}
which is computed by the distributed NOA Algorithm (DNOA) \ref{alg3}.
\begin{algorithm}[!t]\caption{DNOA \ref{alg3} for solving (\ref{HEV3})}\label{alg3}
  \begin{algorithmic}[1]
\State {\bf Set} $\kappa = 0$ and $W^{(0)}(t)=\hat{W}(t)$, where $\hat{W}(t)$ is found by solving (\ref{HEV5}).
  \State {\bf Until} ${\mbox{Trace}}(W^{(\kappa)}(t))- (w_{\max}^{(\kappa)}(t))^H W(t) w_{\max}^{(\kappa)}(t) \leq \epsilon$
{\bf solve} $\{\ds\min_{W(t), R(t)} \ds \sum_{k\in {\cal G}}f(P_{{g_k}})\delta_t+ \mu(\Tr(W(t))-
(w_{\max}^{(\kappa)}(t))^H W(t) w_{\max}^{(\kappa)}(t))$ $\mbox{s.t.}$ (\ref{HEV2d})-(\ref{HEV2b})$\}$
to generate $W^{(\kappa+1)}(t)$ and $R(t)$, and
{\bf reset} $\kappa+1\rightarrow \kappa$.
   \State {\bf Accept} $W^{(\kappa)}(t)$ and $R(t)$  as a found solution of (\ref{HEV3}).
   \end{algorithmic}
\end{algorithm}
\section{Simulation results}
\subsection{Simulation setup}
The SDPs (\ref{HORr}), (\ref{rW5}), (\ref{HEV5}) and (\ref{W5}) are computed using Sedumi\cite{S98} interfaced by CVX \cite{cvx}
on a Core i5-3470 processor. Four power networks from Matpower \cite{ZMT11} are chosen.
The tolerance $\epsilon = 10^{-4}$ is set for the stop criterions.

\begin{figure}[h]
\centering
\includegraphics[width=0.8 \columnwidth]{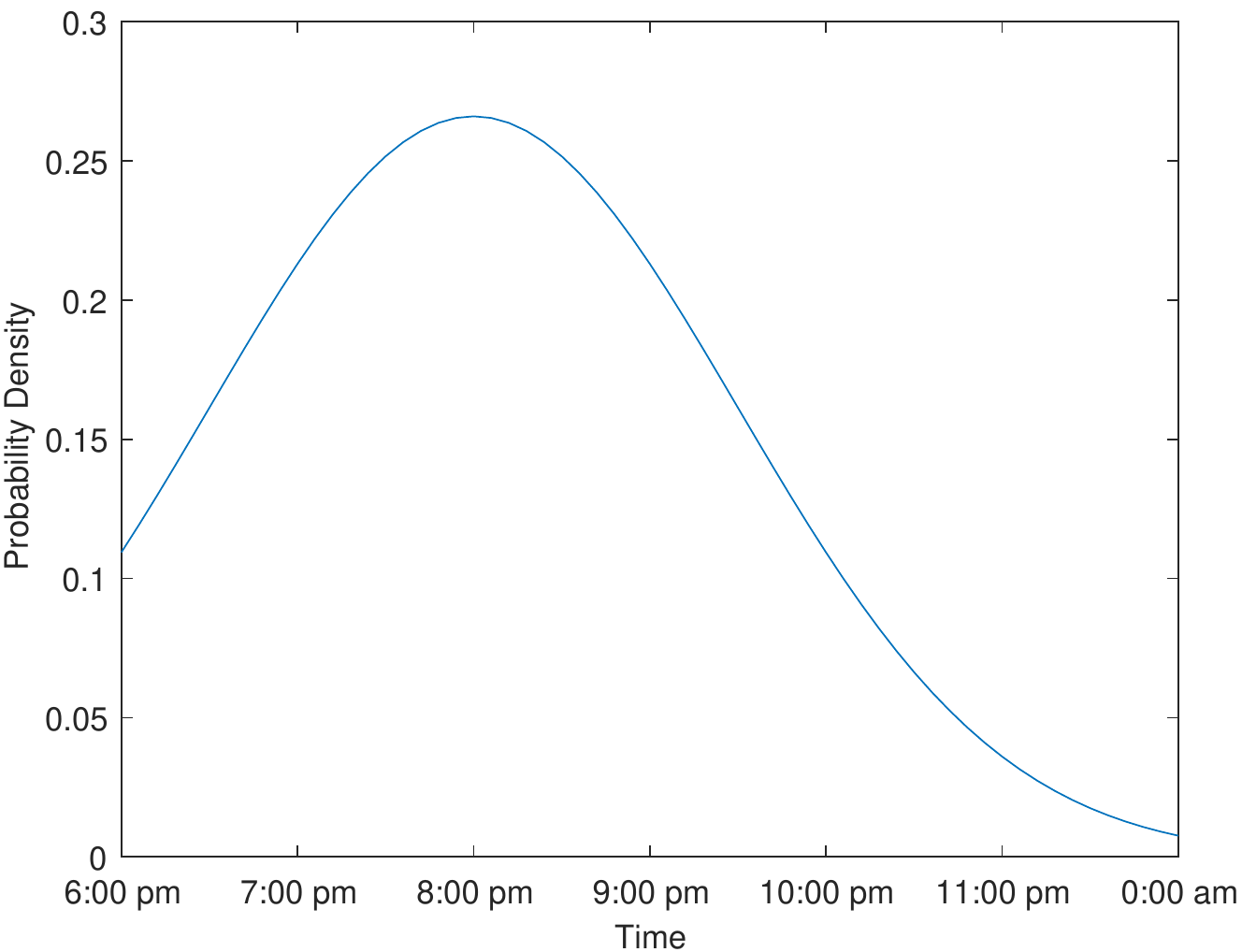}
\caption{The probability density of PEVs' arrivals}
\label{n_arrival}
\end{figure}

Generally, PEVs are charged  after their owners' working hours.
We focus on the charging period from 6:00 pm to 6:00 am of the next day, which is then uniformly
divided into $24$ time slots of $30$ minute length  \cite{JTG13}. Accordingly, the charging time horizon
 is $\clT=\{1, 2, \dots, 24\}$. It is also reasonable to assume that the PEVs arrive during the time period
 from 6:00 pm to midnight. The PEVs must be fully charged after being plugged into the grid. The arrival times of
 PEVs are assumed to be independent and are generated by a truncated  normal distribution $(20,1.5^2)$, which is depicted by Fig. \ref{n_arrival}.

We assume that the PEVs are Tesla Model S's, which have a battery capacity of  100 KWh \cite{Tesla}.
The SOC of all PEVs is set as 20\%. The structure and physical limits of the considered grids are given in the Matpower
library \cite{ZMT11} together with the specific cost functions $f(P_{{g_k}}(t))$.

 Without loss of generality,  PEV loads are connected at the generator buses, which means each generator bus will serve as a charging station.

\begin{figure}[h]
\centering
\includegraphics[width=0.8 \columnwidth]{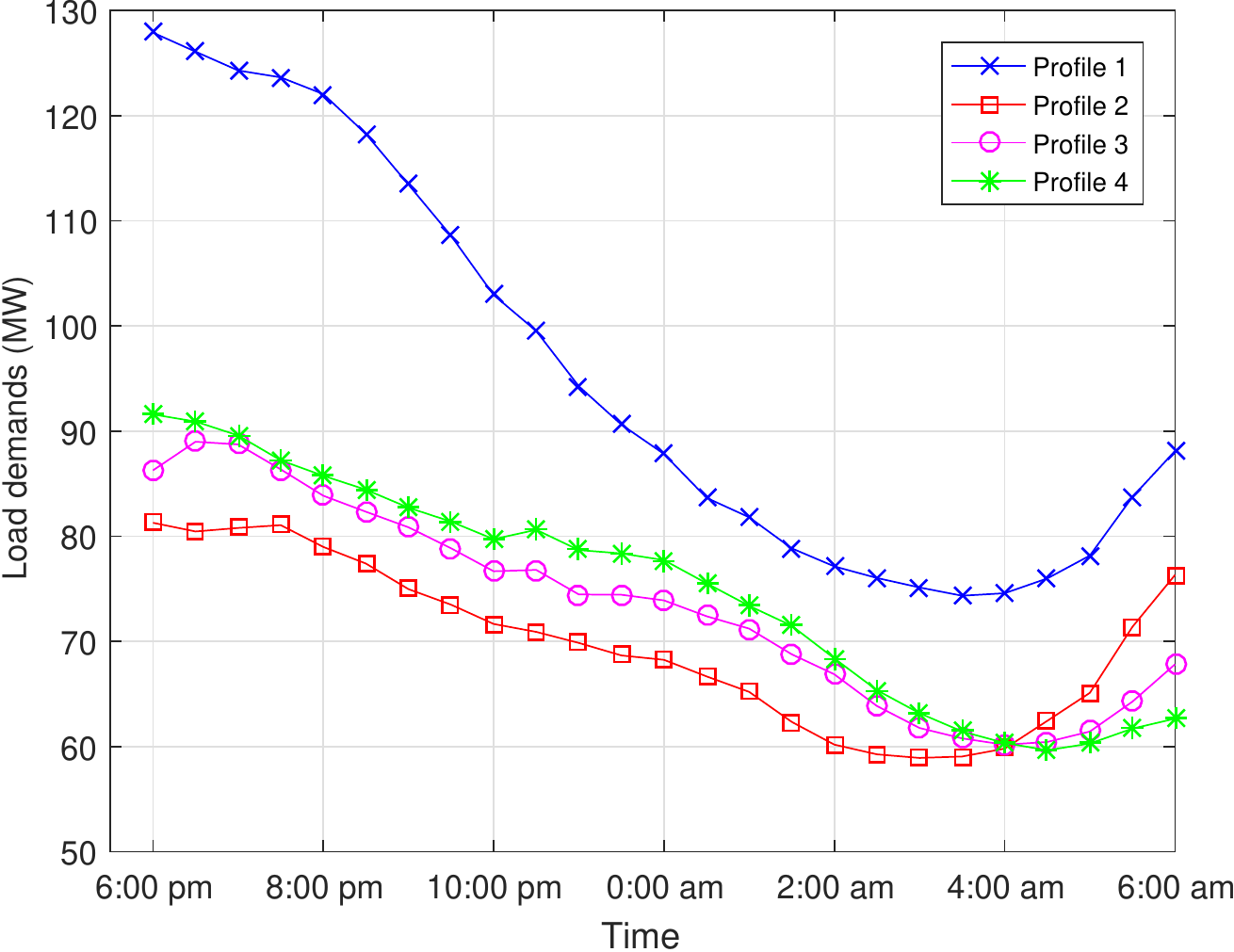}
\caption{Residential load demands of four profiles}
\label{Load_demand}
\end{figure}

\begin{figure}[h]
\centering
\includegraphics[width=0.8 \columnwidth]{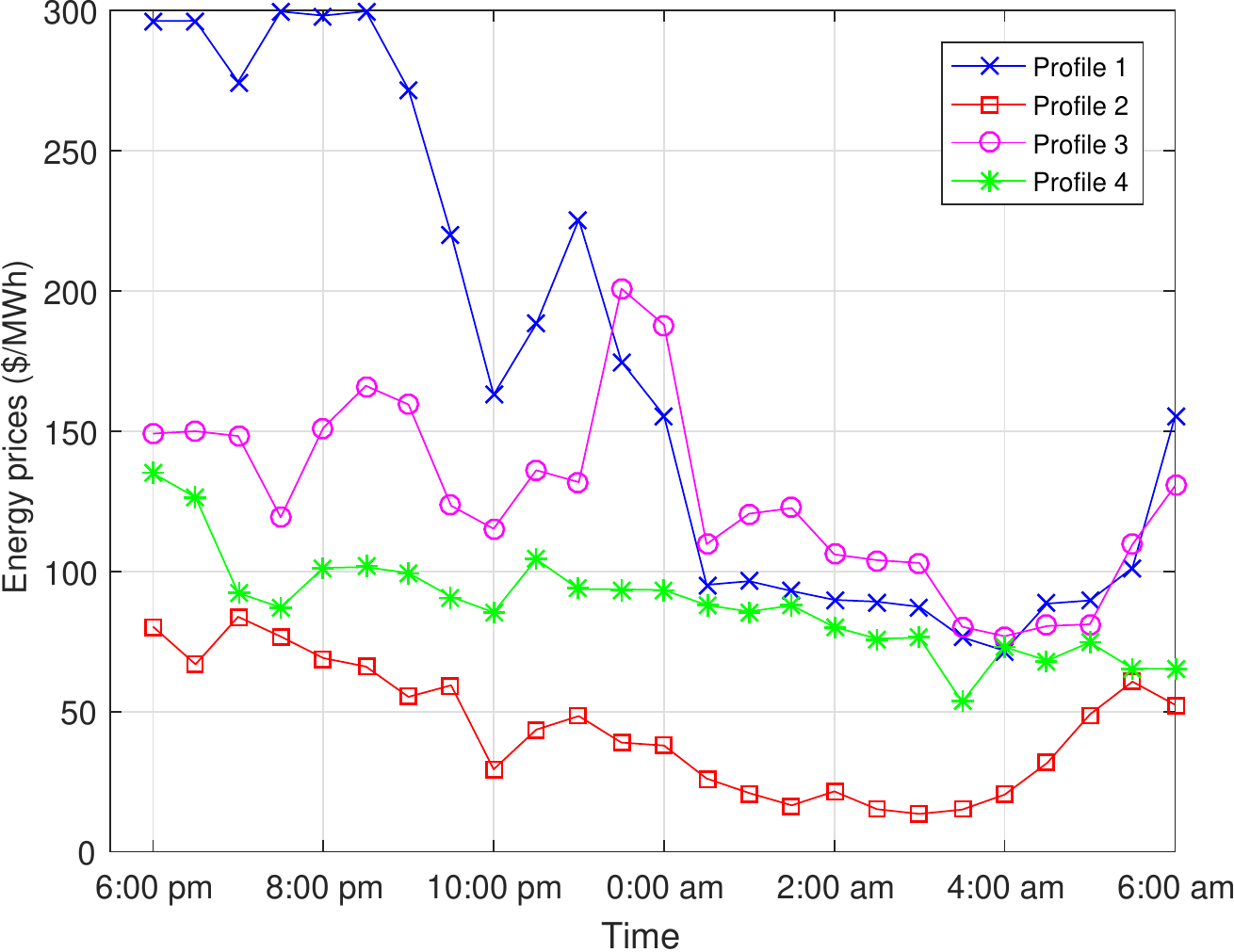}
\caption{Energy prices for four profiles}
\label{Energy_price}
\end{figure}

The price-inelastic load $P_{l_k}(t)$ is calculated as
\begin{equation}
P_{l_k}(t) = \frac{l(t)\times \bar{P}_{l_k}\times T}{\sum_{t=1}^{24} l(t)}, \quad t \in \clT,
\end{equation}
where $\bar{P}_{l_k}$ is the load demand specified by \cite{ZMT11} and $l(t)$ is the residential load demand
taken from \cite{NSW_data}. Four profiles are taken from different days in 2017. Profile 1 is the residential load and energy price from 6:00 pm on February 5th to 6:00 am on February 6th, Profile 2 is from 6:00 pm on March 5th to 6:00 am on March 6th, Profile 3 is from 6:00 pm on April 5th to 6:00 am on April 6th, and Profile 4 is from 6:00 pm on May 5th to 6:00 am on May 6th. Fig. \ref{Load_demand} and Fig. \ref{Energy_price} provide the residential load demand and energy price for these  profiles.

\subsection{MPC-based online computational results}

\subsubsection{Four network simulation}
We test MPC-based online computation for  Case9, Case14, Case30 and Case118mod from \cite{ZMT11} and profile 2
of the residential data. The information on these networks is given  in Table. \ref{Case_inform}, where the first column is the name of network,
the second column  indicates the numbers of buses, generators and branches.
The dimension of $W(t)$ is given in the third column, while the total number of PEVs is shown in the last column.
\begin{table}[h]
    \centering
    \caption{Information on four networks}
    \begin{tabular}{cccc}
    \hline
    & Buses/Generators/Branches & Dim. of $W(t)$ & PEVs \\
    \hline
    Case9 & 9/3/9  & $\mathbb{C}^{9\times 9}$ & 291\\
    Case14& 14/5/20 & $\mathbb{C}^{14\times 14}$ & 485\\
    Case30& 30/6/24 & $\mathbb{C}^{30\times 30}$ & 582\\
    Case118mod& 118/54/186 & $\mathbb{C}^{118\times 118}$ & 5238\\
    \hline
    \end{tabular}
\label{Case_inform}
\end{table}
The computational results are summarized in Table \ref{mpc_network}.
\begin{table}[h]
    \centering
    \caption{MPC results}
    \begin{tabular}{ccccccc}
    \hline
     &Rank &$\mu$&LB &Comp.value& Time(s) \\
    \hline
    Case9& 9 &10 & 27991.4&27992.3&  7.4 \\
    Case14& 1 &- & 40824.1&40824.1& 8.5 \\
    Case30& 1 &- & 4935.6&4935.6&  8.7 \\
    Case118mod& 2 &100 & 644245.9& 644278.5 & 432.1\\
    \hline
    \end{tabular}
\label{mpc_network}
\end{table}
Again, the first column is the network name. The second column presents the initial rank of
the optimal solution  $\hat{W}(t)$ of SDR (\ref{HORr}).
It is observed that the  rank of $\hat{W}(t)$ is the same for all $t\in \clT$. The value of
the penalty parameter $\mu$ in (\ref{rW3}) is given in the third column. As the initial rank of Case14 and Case30 are all rank-one, SDR (\ref{HORr}) already outputs the optimal solution for  (\ref{HOR1}). Comparing the lower bound (LB) in
the fourth column by solving SDR (\ref{HEV5}) at each time and the value found by the proposed MPC-based computation with using NOA  \ref{alg1} in computing (\ref{HEV1}) in each time
reveals the capability of  the MPC-based computation for (\ref{HEV1}).
These values are either the same (for  Case14 and Case30) or almost the same (for Case9 and Case118mod),
so indeed the proposed MPC-based computation could exactly locate a globally optimal solution.
The average  running time for solving (\ref{HOR1}) to implement the proposed MPC-based computation
is provided in the sixth column, which is very short compared with the $30$ minute time slot and thus is
practical for this particular online application.

\begin{figure}[h]
\centering
\includegraphics[width=1.0 \columnwidth]{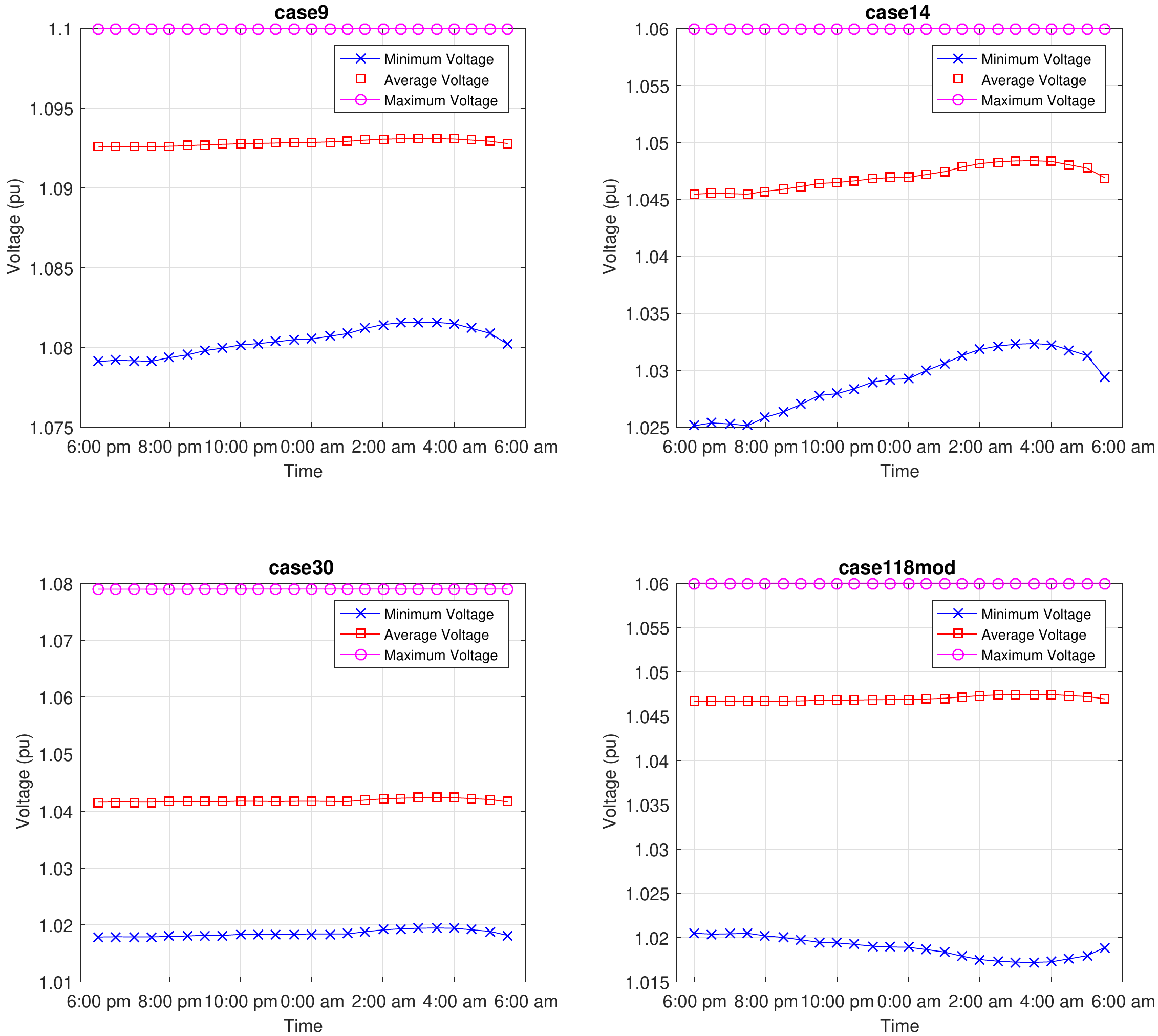}
\caption{Voltage profile for four networks during the charging period}
\label{Voltage}
\end{figure}

The voltage profile for the four networks during the charging period are shown in Fig. \ref{Voltage}. For all cases, the voltage  bound constraints (\ref{HEV1f}) are satisfied. The voltage behavior is stable and smooth.

\subsubsection{Four residential profile simulation}
 We consider Case30 together with four different residential profiles.
 The computational results are provided in Table \ref{mpc_load}, whose format is similar to Table \ref{mpc_network}.
\begin{table}[h]
    \centering
    \caption{MPC results for Case30 with four different residential profiles}
    \begin{tabular}{ccccccc}
    \hline
     &Rank &$\mu$ &LB &Comp. value &  Time(s)\\
    \hline
    Profile 1& 2 &10 & 31961.2&31963.5 & 10.9 \\
    Profile 2& 1 &- & 4963.3&4963.3 & 8.7  \\
    Profile 3& 2 &10 & 10771.3&10774.7  & 8.7  \\
    Profile 4& 1 &- & 8139.3&8139.3& 8.1  \\
    \hline
    \end{tabular}
\label{mpc_load}
\end{table}
It can be seen that, even for the same network, the  rank of the optimal solution $\hat{W}(t)$ of SDR (\ref{HORr})
may be different depending on the residential profiles. For profile 2 and profile 4, the initial rank is one and
SDR (\ref{HORr}) has located a globally optimal solution. However, for profile 1 and profile 3, NOA \ref{alg1} is needed for obtaining the rank-one solution. The convergence speed is fast, and the optimum values are all equal or close to the lower bound, which clearly shows the global efficiency of the proposed MPC-based computation.

\begin{figure}[h]
\centering
\includegraphics[width=0.8 \columnwidth]{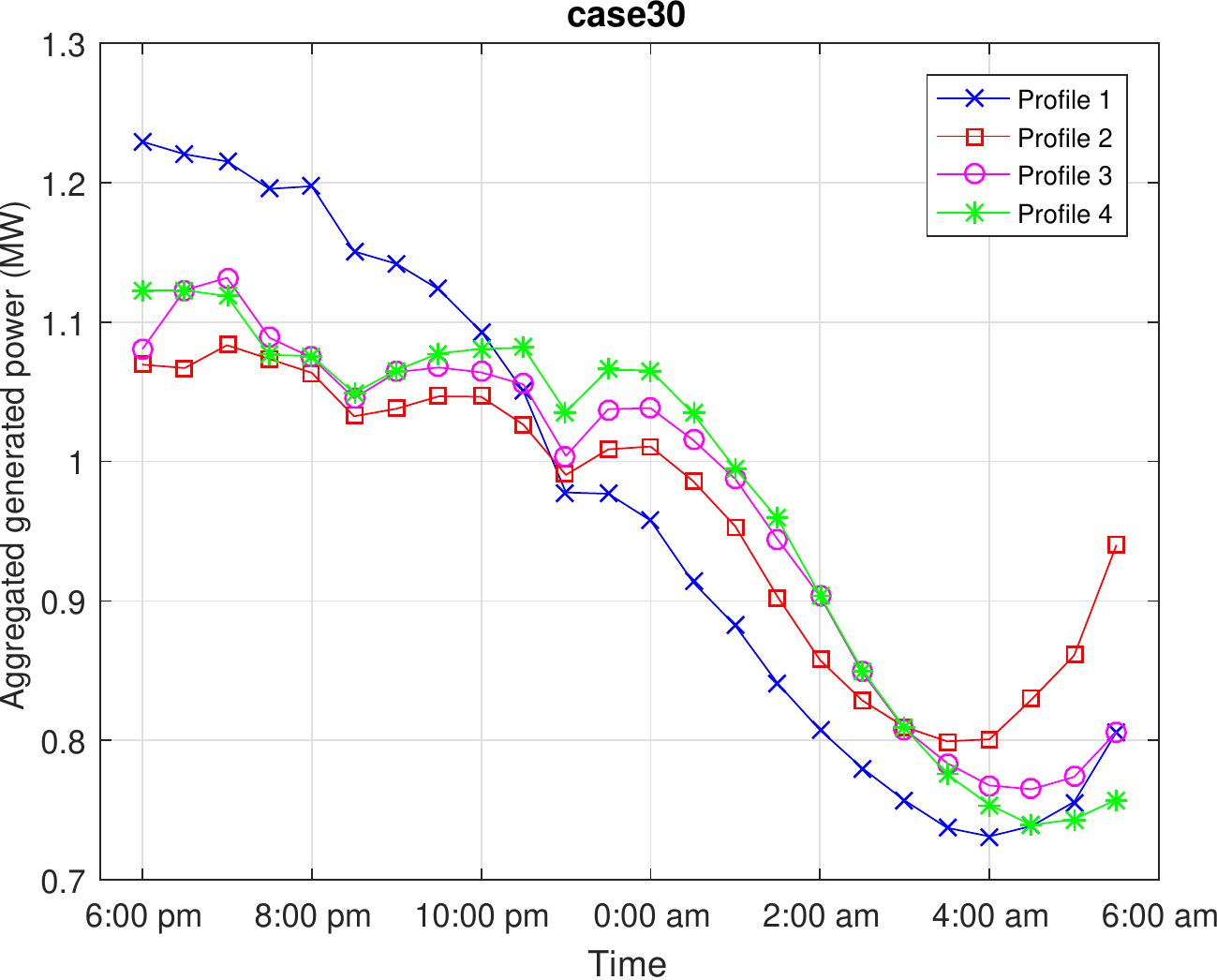}
\caption{Aggregated active power of online charging for Case30 under four residential profiles}
\label{Case30_power_4_profiles_online}
\end{figure}

The aggregated active powers generated at each time are shown in Fig. \ref{Case30_power_4_profiles_online}, from which  the trends of generated power are seen to be similar to the residential load demand in Fig. \ref{Load_demand}.

\begin{figure}[h]
\centering
\includegraphics[width=1.0 \columnwidth]{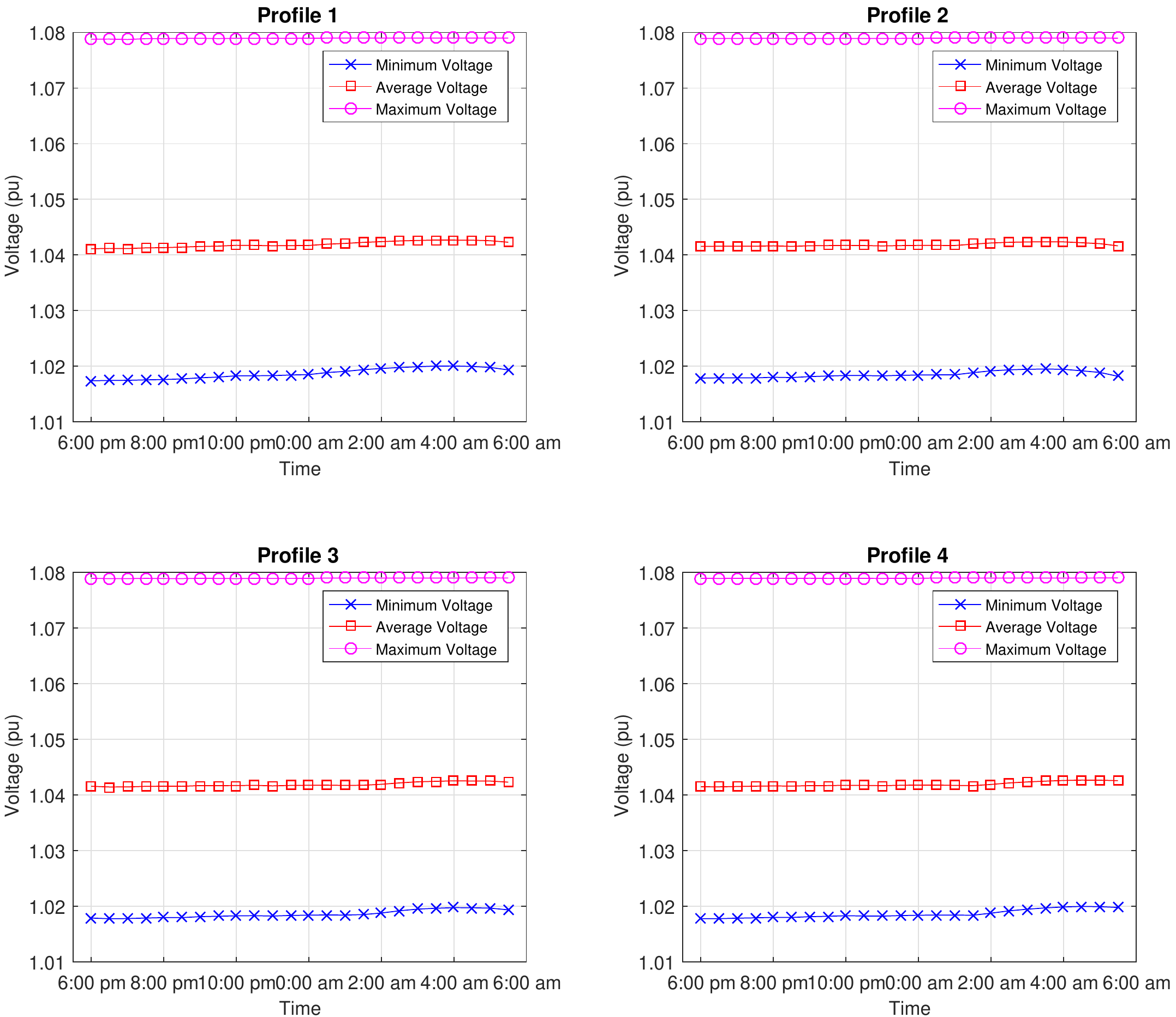}
\caption{Voltage profile of online charging for Case30 under four residential profiles}
\label{Case30_voltage_4_profiles_online}
\end{figure}

The stable and smooth voltage profile for these 4 residential profiles during the charging period are shown in Fig. \ref{Case30_voltage_4_profiles_online}.

\subsection{Off-line computation and comparison with MPC-based online computation}
Firstly,  Case9, Case14, Case30 and Case118mod are tested with the residential data of profile 2 to analyze the efficiency of off-line computation by using Algorithm \ref{alg2} and Algorithm \ref{alg3}. The computational results are summarized in Table \ref{offline_network}.
\begin{table*}[!t]
    \centering
    \caption{Offline results of optimal PEV charging for four networks}
    \begin{tabular}{ccccccccc}
    \hline
     &Rank&$\mu$ &Lower bound&Computed value &Opt. degree &Iterations&NOA time(s)&DNOA time(s)\\
    \hline
    Case9& 9 &1 & 27978.1&27978.1& 100\% &2 & 11.2 & 23.2\\
    Case14& 1 &- & 40800.7&40800.7& 100\% &1 & 8.9 & 8.9 \\
    Case30& 1 &- & 4935.6&4935.6& 100\% &1 & 24.5 & 36.3\\
    Case118mod& 2 &50 & 644225.3&644233.9& 99.999\%&3 & 1094.8 & 363.5\\
    \hline
    \end{tabular}
\label{offline_network}
\end{table*}
The  initial rank in the second column is the rank of the optimal solution $\hat{W}(t)$ of SDR (\ref{HEV5}),
which  is the same for all $t\in \clT$. The value of penalty parameter $\mu$ in (\ref{W5}) is in the third column.
 The fourth column provides the lower bound by computing SDR (\ref{HEV5}). The values found by solving (\ref{W3}) and
(\ref{HEV3})  by Algorithm \ref{alg2} and Algorithm \ref{alg3}  are in the fifth column as they are the same and are either
exactly same as their lower bounds in the fourth column (Case9, Case14 and Case30) or almost the same (Case118mod).
According to the seventh column both Algorithm \ref{alg2} and Algorithm \ref{alg3} converge in two and three iterations
for Case9 and Case188mod, while for Case9 and Case30, SDR (\ref{HEV5}) already outputs the optimal rank-one solution.
 The running times of  Algorithm \ref{alg2} and Algorithm \ref{alg3} are provided in the eighth and ninth column, respectively.  Algorithm \ref{alg2}
requires less running time for small-scale networks such as Case9, Case14 and Case118mod. However, its running time
increases dramatically for large-scale networks  such as Case118mod, for which the scalable  Algorithm \ref{alg3} is
clearly advantageous.

 A performance comparison between MPC-based computation and off-line computation for Case9 and Case30 with the four mentioned residential
profiles is provided in Table \ref{online_offline}, which clearly shows the global optimality of the proposed MPC-based
computation as it attains objective values very close to the lower bounds provided by the off-line computation.
\begin{table*}[!t]
    \centering
    \caption{Performance comparison under MPC-based and off-line computations}
    \begin{tabular}{cccccccc}
    \hline
    & &Rank  & MPC &Offline&Offline/MPC &MPC times(s)&Offline times(s)\\
    \hline
    \multirow{4}{*}{Case9} & Profile 1& 9  & 31963.5&31963.1& 99.99\% & 161.1 & 15.2\\
    &Profile 2& 9  & 27992.3&27978.1& 99.94\% & 177.4 & 15.1\\
    &Profile 3& 9  & 31102.9&30885.1& 99.29\% & 173.7 & 14.8\\
    &Profile 4& 9  & 29896.2&29870.9& 99.91\% & 178.4 & 15.0\\\hline
    \multirow{4}{*}{Case30} & Profile 1& 2  & 31963.5&31963.1& 99.99\%  & 262.4 & 37.4\\
    &Profile 2& 1  & 4963.3&4935.6& 99.43\% & 209.8 & 12.6\\
    &Profile 3& 2  & 10774.7 &10330.8& 95.70\% & 208.2 & 24.5\\
    &Profile 4& 1  & 8139.3&8087.2& 99.35\% & 194.9 & 12.9\\
    \hline
    \end{tabular}
\label{online_offline}
\end{table*}

\begin{figure}[h]
\centering
\includegraphics[width=1.0 \columnwidth]{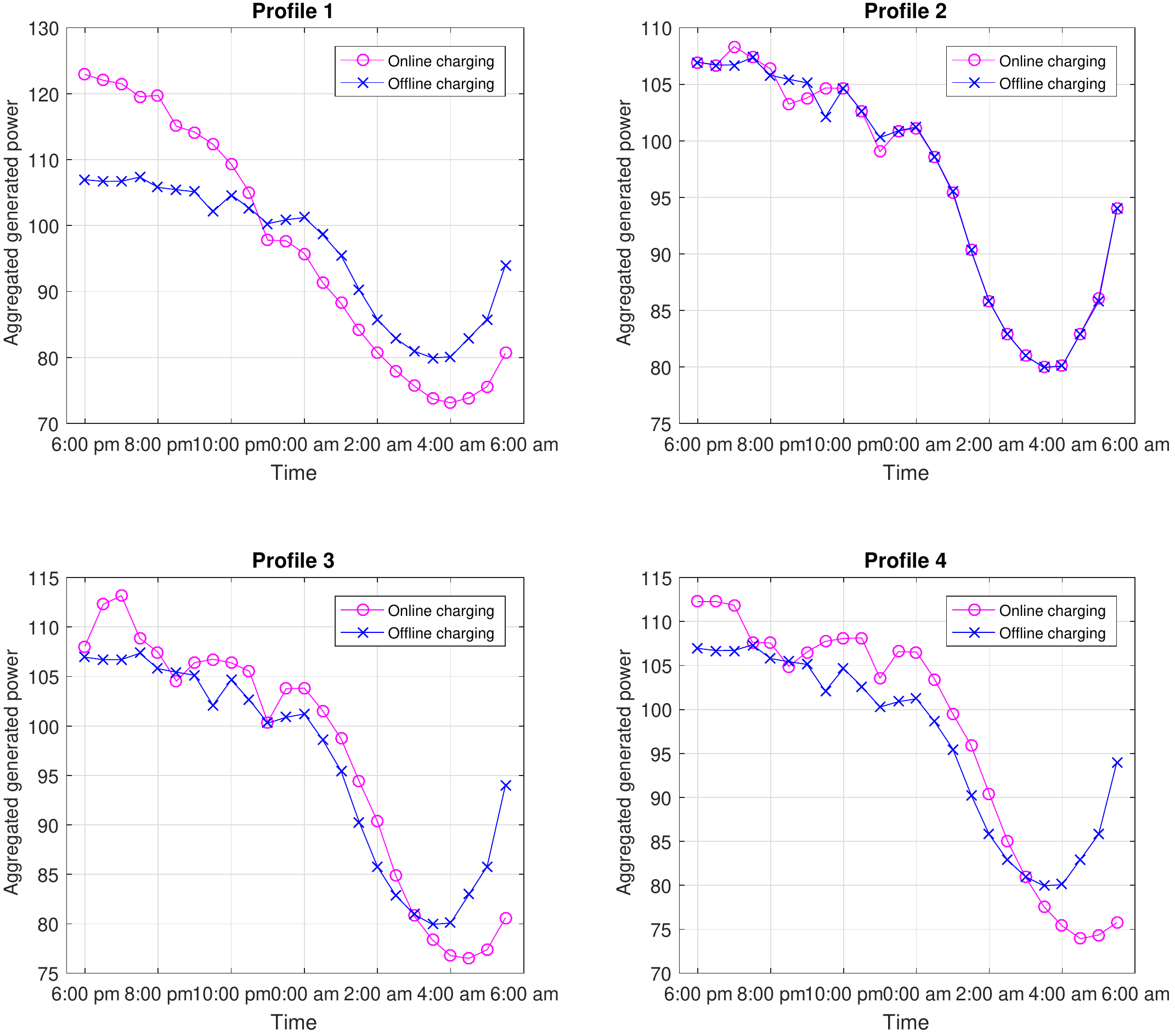}
\caption{Power generation under  MPC-based (online)  and offline computation for  Case30 with four residential profiles}
\label{power_online_offline_Case30}
\end{figure}

\begin{figure}[h]
\centering
\includegraphics[width=1.0 \columnwidth]{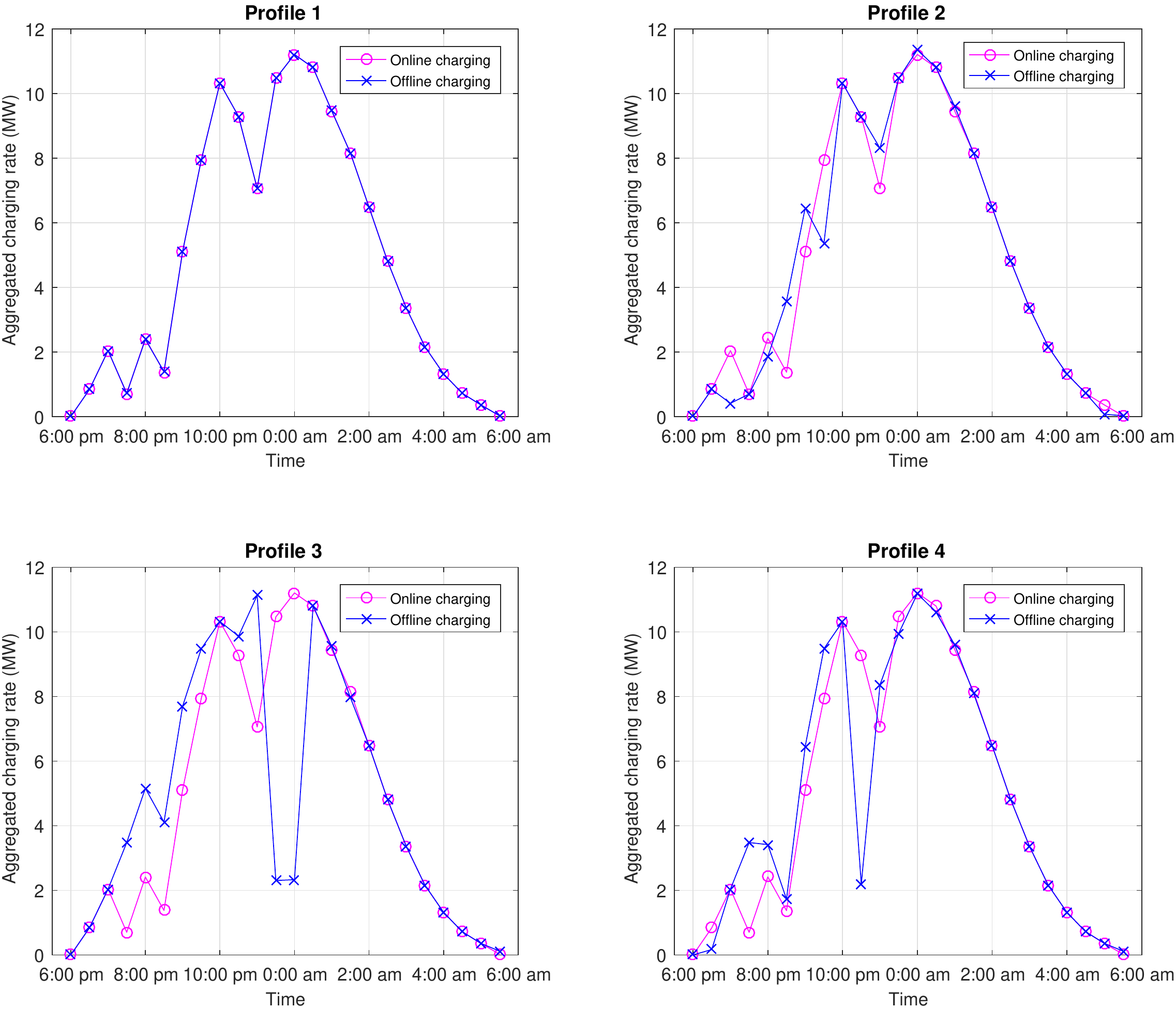}
\caption{PEVs charging load under  MPC-based and offline computation for Case30 with four residential profiles}
\label{charging_online_offline_Case30}
\end{figure}

Fig. \ref{power_online_offline_Case30} plots online and offline power generations in
Case30 with four residential profiles, while Fig. \ref{charging_online_offline_Case30} plots
the corresponding PEV charging scheduling.  The charging load drops dramatically after $0:00$ am, by which all PEVs have been integrated  into the grid but some of them have already been fully charged. Obviously, the charging load
is sensitive to the energy price. For example in profile 3, the increase of the energy price at $11:30$ pm and $0:00$ am leads to a significant drop of the charging load. The charging load under MPC-based and off-line simulation are the same after $0:00$ am because there are no new PEVs arriving  after that time.

\section{Conclusions}

Joint PEV charging scheduling and power control for  power grids
to serve both PEVs at a competitive cost  and residential power demands at a competitive operating cost
is very difficult due to the random nature of PEVs' arrivals and  demands. We have proposed a novel
and easily-implemented  MPC-based computational algorithm that can achieve  a globally optimal solution.

\bibliographystyle{ieeetr}
\bibliography{EV_BIB}
\balance
\end{document}